\documentclass[12pt]{article}
\usepackage{amsmath}
\usepackage{amssymb}
\usepackage{amsfonts}

\begin{document}

\begin{center}
{\Large{\bf{Multiboson logical operators, \\ Laughlin states, and Virasoro algebra}}}

\bigskip\bigskip 

{\large{\bf{Francesco Raffa$^{1}$}}, and {\bf{Mario Rasetti$^{\, 2,3}$}}}

\medskip 
$^{1}$ Dipartimento di Meccanica, Politecnico di Torino, \\ 
\smallskip
$^{2}$ Dipartimento di Fisica and Scuola di Dottorato, Politecnico di Torino, \\ 
Corso Duca degli Abruzzi 24; 10129 Torino (Italy) \\  
\smallskip 
$^{3}$ Fondazione ISI - Institute for Scientific Interchange, \\
Viale Settimio Severo, 65; 10133 Torino (Italy) \\
\smallskip 
\end{center}

\begin{abstract}
Leading idea of this manuscript is to discuss the structure and the deep correlations synthesized by the following diagram.

\bigskip\bigskip  
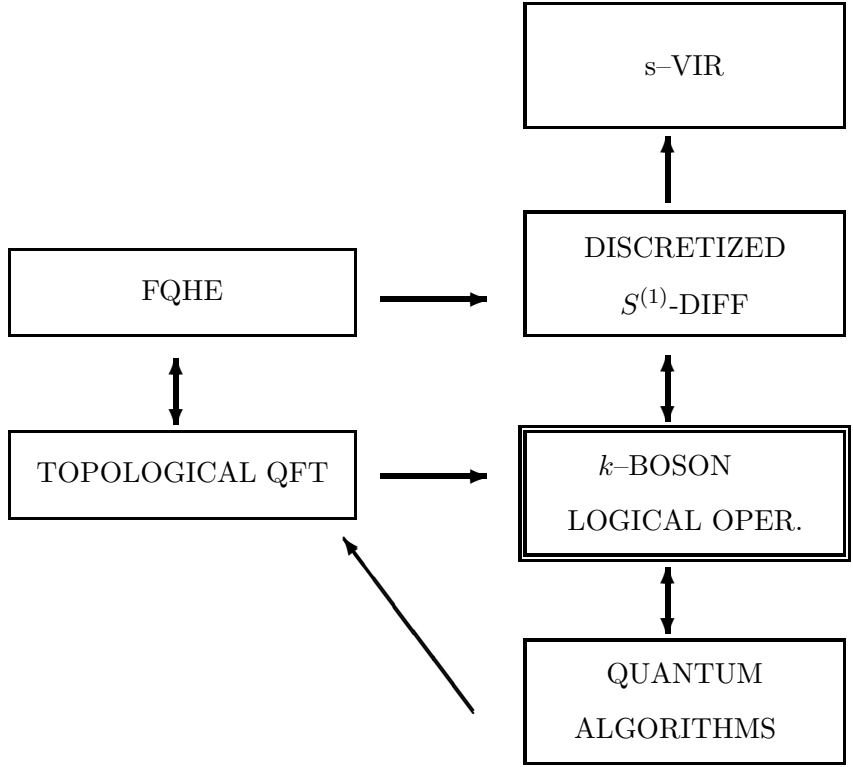
\begin{center} 
\thicklines  
\begin{picture}(250,280)(-94,-135)  
\put(-140,37){\framebox(130,32){FQHE}}  
\put(-140,-32){\framebox(130,32){TOPOLOGICAL QFT}}  
\put(-78,28){\vector(0,-1){25}}
\put(-77,3){\vector(0,1){25}}
\put(-77,28){\vector(0,-1){25}}
\put(-78,3){\vector(0,1){25}}
\put(108,87){\vector(0,1){25}}
\put(109,87){\vector(0,1){25}}
\put(108,29){\vector(0,-1){25}}
\put(0,-16){\vector(1,0){40}} 
\put(0,-17){\vector(1,0){40}} 
\put(109,4){\vector(0,1){25}}
\put(109,29){\vector(0,-1){25}}
\put(108,4){\vector(0,1){25}}
\put(108,-51){\vector(0,-1){25}}
\put(109,-76){\vector(0,1){25}}
\put(109,-51){\vector(0,-1){25}}
\put(108,-76){\vector(0,1){25}}
\put(0,51){\vector(1,0){40}} 
\put(0,50){\vector(1,0){40}} 
\put(35,-105){\vector(-3,4){49}}
\put(35,-106){\vector(-3,4){49}}
\put(55,116){\framebox(120,46){s--VIR}} 
\put(55,37){\framebox(120,46){}} 
\put(115,70){\makebox(0,0){DISCRETIZED}}
\put(115,50){\makebox(0,0){$S^{(1)}$-DIFF}}
\put(55,-46){\framebox(120,46){}}
\put(53,-48){\framebox(124,50){}}
\put(67,-16){\makebox{$\;\;\;\;\;k$--BOSON}} 
\put(67,-37){\makebox{ LOGICAL OPER.}} 
\put(55,-125){\framebox(120,46){}} 
\put(115,-92){\makebox(0,0){QUANTUM}} 
\put(105,-112){\makebox(0,0){$\;\;\;$ ALGORITHMS}} 
\end{picture}  
\end{center} 

In particular, the role of logical operators constructed out of multi-boson (and possibly fractionary-boson) operators will be analyzed in 
its relation on the one hand with the algebra of diffeomorphisms of the circle, on the other with the physical properties of the fractional  
quantum Hall effect (described in terms of Laughlin states) and of its many electron representation in terms of the Chern-Simons topological 
quantum field theory.  

\end{abstract}

\section{Ground state of many--fermion systems}

In the fractional quantum Hall effect (FQHE) (assumed in this manuscript not only as a possible implementation setting, but as a general 
metaphor), as well as in other contexts, such as two-dimensional one-component plasma or statistical mechanics problems in ensembles of random 
matrices, one encounters (non-relativistic, infinitely) degenerate energy levels. In the FQHE case such levels describe charged particles 
constrained in a transverse plane by the interaction with a magnetic field and the levels are Landau levels, whose filling fraction labels the 
observed plateaux in the Hall conductance. 

The basic idea is here that the two-dimensional configuration space of charged particles of the Hall effect should be equivalent to a 
one-degree-of-freedom phase space, quantum and therefore non-commutative \cite{DuJa}. In the presence of a (strong) uniform magnetic field 
${\bf{B}} = B {\bf{e}}_z$, the Lagrangian reads 
\begin{equation}   
{\cal{L}} = \frac{1}{2}\, m {\dot{\bf{x}}}^2 + \frac{q}{c} {\bf{A}}\cdot{\dot{\bf{x}}} - V({\bf{x}}) \; , \label{Lagr}
\end{equation}
where the weak confining potential $V({\bf{x}})$ can, for simplicity but with no essential loss of generality, be neglected  
whereas, adopting the symmetric radial gauge, the connection ${\bf{A}}= \frac{1}{2} \, {\bf{B}} \times {\bf{x}}$ has components $A_i = 
\frac{1}{2} B \epsilon_{i,j} x_j$. There follows that (conserved) Hamiltonian, angular momentum, and momentum can be thought of, 
respectively, as 
\begin{equation}
H = \frac{1}{2}\, m {\dot{\bf{x}}}^2 \; , \; L \equiv L_z = m \, \left ( {\bf{x}} \times {\dot{\bf{x}}} \right ) \cdot {\bf{e}}_z 
+ \frac{1}{2} \, m \omega \, {\bf{x}}^2 \; , \;  {\bf{p}} = m {\dot{\bf{x}}} + \frac{q}{c} \, {\bf{A}} \; , \nonumber 
\end{equation}
where $q$ denotes the charge (assumed positive) and $\omega = {q B}/{m c}$ is the cyclotron frequency. In terms 
of $'$center of mass$\, '$ and relative coordinates, 
\begin{equation}
{\bf{X}} \doteq {\bf{x}} - {\bf{r}} \equiv ( X , Y ) \; , \; {\bf{r}} \doteq \omega^{-1} \, {\dot{\bf{x}}} \times {\bf{e}}_z  \equiv 
( x_r , y_r ) \; , \nonumber 
\end{equation}  
one can define two commuting harmonic oscillator algebras $h(1)$, generated by bosonic operators $\{ a , a^{\dagger} , 
{\mathbb{I}} \}$ and $\{ b , b^{\dagger} , {\mathbb{I}} \}$ 
\begin{equation}
a \doteq \frac{1}{{\sqrt{2}} \lambda_B} \, \left ( X - i Y \right ) \; , \; b \doteq \frac{1}{{\sqrt{2}} \lambda_B} \, \left ( x_r + i y_r 
\right ) \; , \nonumber 
\end{equation}
with $\displaystyle{\lambda_B \equiv {\sqrt{\frac{\hbar c}{q B}}}}$ the $'$magnetic length$\, '$, which are such that 
\begin{equation}
H = \hbar \omega \left ( b^{\dagger} b + \frac{1}{2} \right ) \; , \; L = \hbar \left ( a^{\dagger} a - b^{\dagger} b \right ) \; . \nonumber 
\end{equation}
In the eigenspace ${\cal{H}} = {\rm{span}} \{ |\ell , n \rangle \, |\, \ell , n \in {\mathbb{N}} \}$ of the two-dimensional harmonic oscillator  
(where, due to the cylindrical symmetry, 
\begin{equation}
\langle {\bf{r}} \, |\, n\rangle \equiv \psi_{n , \ell} ( r , \vartheta ) = {\cal{N}}^{-\frac{1}{2}} \, \left ( \frac{r}{\lambda_b} 
\right )^{\ell} {\rm{e}}^{i \ell \vartheta} \, L_{n}^{( \ell )} \left ( \frac{r}{\lambda_b} \right ) \, \exp \left [ - \frac{1}{2} 
\left ( \frac{r}{\lambda_b} \right )^2 \right ] \; , \nonumber 
\end{equation} 
$\displaystyle{L_{n}^{( \ell )}}$ denoting the generalized Laguerre polynomials; in the dimensionless complex coordinate $\displaystyle{ z\doteq 
\frac{r}{\lambda_B} \, {\rm{e}}^{i \vartheta}}$, $\psi_{n , \ell} ( z ) = {\cal{N}}^{-\frac{1}{2}} \, \left ( z \right )^{\ell} \, L_{n}^{( 
\ell )} \left ( |z| \right ) \, \exp \left [ - \frac{1}{2} |z|^2 \right ]$),   
the energy eigenvalues are $E_n = \hbar \omega ( n + \frac{1}{2} )$, whereas the angular momentum eigenvalues have value $L_{n,\ell} = \hbar ( 
\ell - n)$. The quantum number $n$ specifies what is referred to as a Landau level, and for each specific Landau level the fast relative coordinates 
are $'$frozen out$\, '$ and the resulting system has a one dimensional dynamics. This holds in particular for the ground state ($n=0$), where 
the only significan quantum number left is $\ell$.  

On the other hand, varying the magnetic field results in a circular electric field, related to ${\bf{B}}$ by the Maxwell equation 
\begin{equation}
{\rm{rot}} {\bf{E}} = - \frac{1}{c} \,\frac{\partial{\bf{B}}}{\partial t} \; . \nonumber 
\end{equation}  
This in turn induces a change in the angular momentum, because the circular electric field generates a torque on the particle; indeed, since  
\begin{equation}  
\int_S {\rm{rot}} {\bf{E}} \cdot {\rm{d}}^2 S = \int_{\partial S} {\bf{E}} \cdot {\rm{d}}{\bf{\ell}} = E 2\pi r = \frac{1}{c} \,\frac{{\rm{d}}}{{\rm{d}}t} 
\int_S {\bf{B}} \cdot {\rm{d}}^2 S = - \frac{1}{c} \, {\dot{\Phi}} \; , \nonumber  
\end{equation} 
(where ${\rm{d}}^2 S$ is the area element over the surface $S$ whose boundary $\partial S$ is the circle of radius $r$, field line of ${\bf{E}}$, 
whereas ${\rm{d}}{\bf{\ell}}$ is the line element along $\partial S$, and $\Phi$ is the magnetic flux through $S$), and the torque is $M_z = q E r$, 
one has 
\begin{equation} 
\frac{{\rm{d}}L_z}{{\rm{d}}t} = - \frac{q {\dot{\Phi}}}{2\pi c} \; . \nonumber
\end{equation}
There results an anyonic behaviour. Indeed, considering for simplicity a two-particle system, with Hamiltonian 
\begin{equation}
H = \frac{1}{2m} \left [ \left ( {\bf{p}}_1 - \frac{q}{c} {\bf{A}}_1 \right )^2 + \left ( {\bf{p}}_2 - \frac{q}{c} {\bf{A}}_2 \right )^2 \right ] \; , 
\nonumber 
\end{equation}
where 
\begin{equation}
{\bf{A}}_{1,2} = \pm \frac{\Phi}{2\pi}\, {\bf{e}}_z \times \frac{{\bf{r}}}{r^2} \; , \; {\bf{r}} \doteq {\bf{x}}_1 - {\bf{x}}_2 \; , 
\; r = | {\bf{r}} | \; , \nonumber 
\end{equation}
and writing it in the center-of-mass and relative coordinate system, ${\bf{P}} \doteq {\bf{p}}_1 + {\bf{p}}_2$, ${\bf{p}} \doteq 
\frac{1}{2} \bigl ( {\bf{p}}_1 - {\bf{p}}_2 \bigr )$, 
\begin{equation}
H = \frac{{\bf{P}}^2}{4m} + \frac{1}{m}  \left ( {\bf{p}} - \frac{q}{c} \,\frac{\Phi}{2\pi}\, {\bf{e}}_z \times \frac{{\bf{r}}}{r^2} \right )^2 \; , 
\nonumber
\end{equation}  
results in the decoupled motion of the center-of-mass and that of a charged $'$pseudo$\, '$-particle of mass $\frac{1}{2} m$ moving along an orbit 
that encircles a flux $\Phi$. For ${\bf{r}} \equiv ( r, \vartheta )$, upon performing the gauge transformation 
\begin{equation}
{\bf{A}} \mapsto {\bf{A}} - {\rm{grad}} \left ( \frac{\Phi}{2\pi} \,  \vartheta \right ) \, , \nonumber 
\end{equation} 
(singular, because not single-valued, being $\vartheta$ defined up to integer multiples of $\pi$) in the new gauge the vector potential is 
switched off and $H$ transforms to 
\begin{equation}
H' = \frac{{\bf{P}}^2}{4m} + \frac{{\bf{p}}^2}{m} \; , \nonumber 
\end{equation} 
which is the Hamiltonian of a system of two free particles. The gauge transformation induces a change in the wave function, 
\begin{equation}
\psi ( r,\vartheta ) \mapsto \exp \left ( -i \frac{q \Phi}{c h} \vartheta \right ) \psi ( r,\vartheta ) \; , \nonumber  
\end{equation}
pointing out to an anyonic statistics. The latter is indeed characterized by the feature that the wave function describing the system of two particles 
is multiplied upon particle interchange by the non-trivial phase factor 
\begin{equation}
\alpha = \frac{q \Phi}{2 c \hbar} \; . \label{alpha} 
\end{equation} 

On the other hand, the limit of strong $B$ implies that in (\ref{Lagr}) $m$ is negligible, thus, together with the confining potential, the kinetic energy term can be 
dropped in (\ref{Lagr}). There results that the canonically conjugate momentum of $x_1$ is $p_1 \equiv B x_2$, whereby quantization is achieved (with the 
appropriate scale of units ($\hbar =1$) to simplify notation) by the commutation rule
\begin{equation} 
\left [ x_1 , x_2 \right ] = \frac{i}{B} \; . \label{commut} 
\end{equation}
In the absence of the confining potential all the eigenstates at each Landau level are degenerate, because it is 
$V$ that removes the degeneracy. Single particle eigenstates are those, well known, of the harmonic oscillator, 
\begin{equation} 
|\ell \rangle = \frac{1}{\sqrt{\ell !}}\, {a^{\dagger}}^{\ell} \, |0\rangle \; , \nonumber 
\end{equation}
and the corresponding normalized wave-functions 
\begin{equation}
\langle z \, |\, \ell , n=0 \rangle \equiv \psi_{\ell} (z) = \sqrt{\frac{B}{2\pi \ell !}} \, z^{\ell} \, 
{\rm{e}}^{- \frac{1}{2} |z|^2} \; , \; {\rm{with}} \; z\doteq \sqrt{\frac{1}{2}B}\, (x_1 + i x_2 ) \; . \nonumber 
\end{equation}
It is worth noticing that $\psi_n (z)$ can be thought of as the coherent state representation of $|n\rangle$ over the complex plane ${\mathfrak{C}}$.  

With the appropriate scaling one can assume as complete set of orthogonal states for the lowest Landau level \cite{PrGi} 
\begin{equation}
\psi_{\ell}(z) = \frac{1}{\sqrt{\pi}}\, z^{\ell} \, {\rm{e}}^{-\frac{1}{2}|z|^2} \; , \nonumber 
\end{equation}
where $z$ is a complex variable, and $\ell \in {\mathbb{N}}$, can be interpreted -- in view of the role of $x_2$ -- as the angular momentum 
eigenvalue. The normalization of the wave-functions $\psi_{\ell}$ is 
\begin{equation}
\langle \psi_{\ell}\, |\, \psi_{\ell '}\rangle = \int_{\mathfrak{C}} \frac{{\rm{d}}^2 z}{\pi}\, {\bar{z}}^{\ell} z^{\ell'} \, 
{\rm{e}}^{-|z|^2} = \ell ! \, \delta_{\ell ,\ell'} \; . \nonumber
\end{equation}

For a system of $N$ non-interacting particles with Fermi statistics, since each state can be occupied by at most one fermion, the presence 
of the confining potential selects a unique ground state, which is the minimum angular momentum state. In the lowest Landau level second 
quantized orthogonal basis states $\Psi_{\{ \ell_0 , \ell_1 , \dots , \ell_{N-1} \}}$ $\equiv \Psi_{\{ \ell_0 , \ell_1 , \dots , \ell_{N-1} 
\}} ( z_0 , z_1 , \dots , z_{N-1} )$ are Slater determinants: 
\begin{eqnarray}
 && \Psi_{\{ \ell_0 , \ell_1 , \dots , \ell_{N-1} \}} ( z_0 , z_1 , \dots , z_{N-1} ) = \nonumber \\  
\qquad\; &=& \frac{\pi^{-\frac{1}{2}\, N}}{\sqrt{N !}} \exp\left ( -\frac{1}{2}\, \sum_{i=0}^{N-1} |z_i|^2 \right )  
\cdot {\rm{det}} \, \left | \begin{array}{c c c c} z_0^{\ell_0} & z_0^{\ell_1} & \cdots & z_0^{\ell_{N-1}} \\ 
z_1^{\ell_0} & z_1^{\ell_1} & \cdots & z_1^{\ell_{N-1}} \\
\vdots & \vdots & \vdots & \vdots \\ 
z_{N-1}^{\ell_0} & z_{N-1}^{\ell_1} & \cdots & z_{N-1}^{\ell_{N-1}} \\
 \end{array} \right | \, . \nonumber
\end{eqnarray}
Since the multi-particle wave-function is by construction anti-symmetric with respect to permutations of the $\ell$'s, one can assume, 
with no loss of generality, that sub-indices $\ell_i$ are ordered: $0\leq \ell_0 < \ell_1 < \cdots < \ell_{N-1}$, whence 
\begin{equation}
\langle \Psi_{\{ \ell_0 , \ell_1 , \dots , \ell_{N-1} \}} \, | \, \Psi_{\{ \ell_0' , \ell_1' , \dots , \ell_{N-1}' \}}\rangle = 
\, \prod_{i=0}^{N-1} \ell_i ! \, \delta_{\ell_i , \ell_i'} \; . \nonumber 
\end{equation}
Now, the state with lowest total angular momentum corresponds to the choice $\ell_i \equiv i \, , \, \forall i \, , \, i=0, \dots , N-1$, 
for which the above determinant becomes 
\begin{equation}
\Delta (z_0 , z_1 , \dots , z_{N-1}) \equiv {\rm{det}} \, \left | \begin{array}{c c c c} 1 & z_0 & \cdots & z_0^{N-1} \\ 
1 & z_1 & \cdots & z_1^{N-1} \\
\vdots & \vdots & \vdots & \vdots \\ 
1 & z_{N-1} & \cdots & z_{N-1}^{N-1} \\
\end{array} \right | \, = \, \prod_{i=1}^{N-1}\,\prod_{j=0}^{i-1} \left ( z_i - z_j \right ) \; , \label{Delta}  
\end{equation} 
which is -- up to a factor -- a Vandermonde determinant. Notice that 
\begin{equation}
\int_{{\mathfrak{C}}^{\times N}} \prod_{i=0}^{N-1} \left [ \frac{{\rm{d}}^2 z_i}{\pi}\, {\rm{e}}^{-|z_i |^2} \right ]\, \left | 
\Delta (z_0 , z_1 , \dots , z_{N-1}) \right |^2 = \prod_{j=1}^N j! \; . \nonumber 
\end{equation}
Notice as well that the quantity 
\begin{equation}
{\mathfrak{D}} (z_0 , z_1 , \dots , z_{N-1}) \doteq \prod_{{i,j=0}\atop{( i\neq j )}}^{N-1} \left ( z_i - z_j \right ) \, \equiv \, 
(-)^{\frac{1}{2} N (N-1)} \Delta^2 (z_0 , z_1 , \dots , z_{N-1}) \; , \label{star} 
\end{equation}
is one of the most fundamental objects of algebra: the discriminant of the monic polynomial 
\begin{equation}
{\cal{P}}(z) = \prod_{i=0}^{N-1} \left ( z - z_i \right ) \; , \label{polyn}  
\end{equation}
whose roots are just the $z_i$'s. Recall that the discriminant ${\mathfrak{D}}$ as well as its powers are functions (indeed admit a 
polynomial representation in terms) of the well known invariant elementary symmetric functions ${\mathfrak{s}}_k$, $0 \leq k\leq N$, 
whereby one can write  
\begin{equation}
{\cal{P}}(z) = \sum_{k=0}^N (-)^k {\mathfrak{s}}_k \, z^{N-k} \; , \; {\mathfrak{s}}_0 \equiv 1 \; , \label{sym} 
\end{equation}  
a property well known since the days of Newton. This property will be exploited in what follows viewing it from a particular perspective: 
the ring of symmetric polynomial functions with rational coefficients can be thought of as a graded vector space ${\mathfrak{V}}$ 
over the field of rationals ${\mathbb{Q}}$. Since each homogeneous subspace of ${\mathfrak{V}}$ admits a basis in terms of Jacobi-Schur 
functions $ch_Y (z_i)$, labelled by partitions (or Young tableaux),   
\begin{equation}
{\mathfrak{D}} (z_i)^s = (-)^{\frac{1}{2} s N (N-1)} \, \sum_{|Y|=sN(N-1)} g_Y^{(s)}\, ch_Y (z_i) \, , \nonumber 
\end{equation}
for $s$ any positive integer. The notation $ch$ is to remind us that these are the polynomial characters of the general 
linear group $GL_N$. 

The physical feature that relates FQHE with the quantum Coulomb gas \cite{TSG} is that electrons, even if subject to an intense magnetic 
field, are nevertheless still weakly interacting via Coulomb force (as well as, of course, with the substrate and the unavoidable 
impurities). Understanding the properties of such a system even at very low temperature (ground state) is in itself a quite difficult 
problem, yet the remarkable stability that such quantum fluids exhibit, shown in particular just by the existence of the Hall plateaux, 
led Laughlin to conjecture the existence of a wave function \cite{LAUGH} for filling fraction $\nu = 1/(2s+1)$, $s=0,1,\dots$ (odd 
denominator is required by Fermi statistics) of the form:
\begin{equation}
\phi_s (z_0 , \dots , z_{N-1}) \propto \pi^{-\frac{1}{2} N} \exp\left ( -\frac{1}{2}\, \sum_{i=0}^{N-1} |z_i|^2 \right ) \, \Delta^{2s+1} 
(z_0 , \dots , z_{N-1}) \; , \nonumber 
\end{equation}    
(where the normalization factor was omitted for simplicity) such that as $z_i \rightarrow z_j$ the wave function have zeroes of order 
$2s+1$. 

$\phi_s$ has an expansion on the complete basis of (free) Slater determinants: 
\begin{equation}
\phi_s (z_0 , \dots , z_{N-1}) = \sum_{0\leq \ell_0 < \cdots < \ell_{N-1} \leq (2s+1)(N-1)} g_{\ell_0 , \dots , \ell_{N-1}} 
\Psi_{\{ \ell_0 , \dots , \ell_{N-1} \}} \; , \nonumber 
\end{equation}
or, equivalently, 
\begin{equation}
\Delta (z_0 , \dots , z_{N-1})^{2s+1} = \sum_{0\leq \ell_0 < \cdots < \ell_{N-1} \leq (2s+1)(N-1)} g_{\ell_0 , \dots , \ell_{N-1}} 
\left | z^{\ell_0} \cdots z^{\ell_{N-1}} \right |  \; , \nonumber 
\end{equation}
where the shorthand notation $\displaystyle{\left | z^{\ell_0} \cdots z^{\ell_{N-1}} \right |}$ denotes the determinant of the matrix obtained 
from the given row by replacing $z$ by $z_0$ in the first line, by $z_1$ in the second line, ... , by $z_{N-1}$ in the $N$-th line. This in turn 
amounts to 
\begin{eqnarray}
&&\Delta (z_0 , \dots , z_{N-1})^{2s} = (-)^{\frac{1}{2} \, s N (N-1)} \, {\mathfrak{D}} (z_0 , \dots , z_{N-1})^s \nonumber \\ 
&& \qquad\quad = \sum_{0\leq \ell_0 < \cdots < \ell_{N-1} \leq (2s+1)(N-1)} g_{\ell_0 , \dots , \ell_{N-1}} ch_{\ell_0, \dots , 
\ell_{N-1}} (z_0, \dots , z_{N-1}) \; . \nonumber 
\end{eqnarray}

An algebraic approach to fractional statistics in one dimension was originally proposed in \cite{LeMy}, \cite{HaLeMy}, where it 
was shown that for identical anyons the dynamical algebra does not consist of two (or several) independent $h(1)$ 
algebras, as one might guess from the above analysis, but the separation of coordinates implies that it is a novel 
algebra whose generators are nonlinear in the single particle creation and annihilation operators and whose 
representations are characterized by parameters that can be thought of as depending on statistics. For example, in the  
case of two anyons the resulting algebra is $su(1,1)$, in a representation that can be related to the Holstein-Primakoff 
representation \cite{HP}.    

We argue that, in view of the form of the wave funcion, the resulting algebra for the $N \rightarrow \infty$ particle 
case should have two distinctive features: i) include an (infinite) number of $su(1,1)$ subalgebras, one for each particle; ii) 
be invariant under the group of permutations of the root of polynomial ${\cal{P}}(z)$ or, equivalently, of function 
$\Delta (z_0, z_1, \dots , z_{N-1})$. The latter identifies the group with the holonomy group of the braid group (see 
the seminal work of V.I Arnol'd in \cite{ARN}) and the generating algebra -- in present case where the roots of ${\cal{P}}(z)$ 
are roots of unit -- with that of the diffeomorphisms of the circle. This leads us to conjecture that in this case the relevant 
algebra is Virasoro's algebra. Since the elementary constituents are indeed fermions, we actually propose that the 
relevant algebra should be the ${\mathbb{Z}}_2$-graded (or $'$super$\, '$) Virasoro algebra.  

It is interesting to point out at this point that recently the spin--network quantum simulator model (which 
essentially encodes the quantum deformed $su(2)$ Racah--Wigner tensor algebra) was shown \cite{GaMaRa} to be capable 
of implementing families of finite-states and  discrete-time quantum automata which accept the language generated 
by the braid group. This led to the construction of a quantum algorithm -- indeed based on a discretized version of 
the Chern-Simons topological quantum field theory -- whereby the invariants of 3-manifolds could be efficiently 
evaluated. 

In two spatial dimensions the group relevant to the quantum statistics of particles is the braid group \cite{WU}, \cite{GoMeSh}, 
rather than the permutation group. Indeed the correct configuration space for a system of $N$ indistinguishable particles, if 
${\mathfrak{X}}$ is the single particle configuration space, is not ${\mathfrak{X}}^{\otimes\, N}$, but rather ${\mathfrak{X}}^{\otimes\, 
N}/{\cal{S}}_N$, where ${\cal{S}}_N$ is the (discrete) permutation group on $N$ elements, but the exchange of a generic pair of particles  
in two dimensions, where only one (Cartan) element of the angular momentum algebra survives, allows for the generation of a phase $\alpha$ 
(see (\ref{alpha}) interpolating between $2k\pi$ (bosons), and $(2k+1)\pi \, , \, k\in {\mathbb{Z}}$ (fermions). As a result, the possibility 
for non-standard statistics exists . The latter prospects two cases. Abelian anyons \cite{LeMy-2}, \cite{GoMeSh-anyon}, \cite{Wil} is the first, 
transforming within a unitary abelian representation of the braid group. Anyons in the lowest Landau level, relevant to the quantum Hall effect, 
belong to this set \cite{Hal}, \cite{ArScWi} and provide realizations of ideal exclusion statistics \cite{Hald}, \cite{DaOu}. A natural 
generalization is of course nonabelian anyons and the corresponding statistics, based on nonabelian representations of the braid group. Such 
statistics is the anyonic counterpart of parastatistics \cite{Gre}, \cite{MeGr}. Just as abelian anyons can be thought of as ordinary (bosonic 
or fermionic) particles interacting through a non-dynamical abelian gauge field, nonabelian anyons can be represented as particles carrying 
internal degrees of freedom in some irreducible representation ${\mathcal{R}}$ of nonabelian group $SU(n)$ interacting through an appropriate 
non-dynamical, nonabelian gauge field. What induces the statistics, then, is the group $SU(n)$, the representation ${\mathcal{R}}$ and the 
coupling strength $g$ of the internal degrees of freedom to the gauge field. A field-theoretic approach to achieving such statistics, in 
analogy with the abelian case, is to couple the particles to a nonabelian gauge field with a Chern-Simons action \cite{DeJaTe}, \cite{Ver}, 
letting the system inherit the quantization condition expressed by $g = 2/n$, with integer $n$. 

For $N$ non-interacting spinless particles on the plane with internal degrees of freedom transforming in some finite-dimensional unitary 
irreducible representation ${\mathcal{R}}$ of $SU(n)$, in the gauge where the Hamiltonian of the particles is free, the nontrivial statistics 
manifests in the wavefunction of the system, which is not single valued \cite{IsLoOu}. Under an exchange of particles following a path belonging 
to a given element of the braid group, the wave function transforms as some nonabelian representation of the braid group parametrized by 
the irreducible representation ${\mathcal{R}}$ of $SU(n)$ and the statistics inducing coupling parameter $g$.

\section{The discretized group of diffeomorphisms of quantum $'$phase space$\, '$} 

\subsection{The super Virasoro algebra}

In its most general definition, the Virasoro algebra is a central extension of the complex Witt algebra ${\mathfrak{W}}$ of meromorphic 
vector fields on a genus $0$ Riemann surface that are holomorphic everywhere, except at two fixed points. ${\mathfrak{W}}$, which may be 
defined as the Lie algebra of derivations of complex Laurent functions $\displaystyle{\sum_{k \in {\mathbb{Z}}} a_k z^k}$, where only 
finitely many among the $a_k$'s ($a_k \in {\mathbb{C}}$) are nonzero, can be thought of as well as the complex Lie algebra of real polynomial 
vector fields on the circle, and hence as a dense subalgebra of the Lie algebra of diffeomorphisms of the circle. $\displaystyle{{\mathfrak{W}} 
= \bigoplus_{n \in {\mathbb{Z}}} {\mathbb{C}}\, L_n}$, has basis elements, denoted by $L_n , n \in {\mathbb{Z}}$, given by 
\begin{equation}
L_n \doteq - z^{n+1} \, \frac{{\rm{d}}}{{\rm{d}}z} \; , \; n \in {\mathbb{Z}} \; , \nonumber 
\end{equation}
satisfying the commutation relations 
\begin{equation}
\, \left [ L_m , L_n \right ] = (m-n)\, L_{m+n} \; , \; n,m \in {\mathbb{Z}} \; . \nonumber 
\end{equation}
${\mathfrak{W}}$ contains infinitely many $su(1,1)$ algebras, generated by $\{ L_n , L_{-n} , L_0 \}$, $\forall n \in {\mathbb{N}}\, , \, n>0$.   

${\mathfrak{W}}$ is equipped with a (unique) nontrivial one-dimensional central extension ${\tilde{\mathfrak{W}}} = {\mathfrak{W}} \oplus {\mathbb{C}}{\bar{c}}$  
(up to isomorphisms of Lie algebras) with basis $\{ c \} \cup \{ L_n \, |\, n \in {\mathbb{Z}} \}$, $c \in {\mathbb{C}}{\bar{c}}$, with the 
commutation relations    
\begin{equation}
\left [ c , L_n \right ] = 0 \; ; \; \left [ L_m , L_n \right ] = (m-n) \, L_{m+n} + \delta_{m, -n} \, c \frac{m^3 - m}{12} \; , \; \forall 
n,m \in {\mathbb{Z}} \; . \nonumber 
\end{equation} 
${\tilde{\mathfrak{W}}}$ is called the Virasoro algebra, and denoted by $Vir$. 

There exists an antilinear map $\Omega \, : \, Vir \mapsto Vir$, which is in fact an antilinear anti-involution on $Vir$ ({\sl{i.e.}}, such that 
$\displaystyle{ \left [ \Omega ( L_n ) , \Omega ( L_m ) \right ] = \Omega ( \left [ L_m , L_n \right ] )}$) defined by requiring that $\Omega ( 
L_n ) = L_{-n}$ and $\Omega ( c ) = c$. Contravariance of hermitian forms on representations of $Vir$ and unitarity of such representations are 
always considered with respect to $\Omega$. $Vir$ has a triangular decomposition into the three Lie subalgebras
\begin{equation}
{\mathfrak{n}}_+ = \bigoplus_{n=1}^{\infty} {\mathbb{C}} L_{-n} \quad , \quad {\mathfrak{h}} = {\mathbb{C}}c \oplus {\mathbb{C}} L_0 \quad , 
\quad {\mathfrak{n}}_- = \bigoplus_{n=1}^{\infty} {\mathbb{C}} L_n \; . \nonumber  
\end{equation}
Let $\pi \, : \, Vir \mapsto {\mathfrak{gl}} ({\cal{V}})$ be a representation of $Vir$ in linear space ${\cal{V}}$: ${\cal{V}}$ admits a basis 
consisting of the eigenvectors of $\pi (L_0 )$ which span finite-dimensional eigenspaces and have all their eigenvalues which are non-negative. 

A representation of $Vir$ over ${\cal{V}}$ is said to be a $'$highest weight$\, '$ representation if there exist an element 
${\bf{v}} \in {\cal{V}}$ and two numbers $C, h\in {\mathbb{C}}$ such that $c \, {\bf{v}} = C \, {\bf{v}}$, $L_0 \, {\bf{v}} = h \, {\bf{v}}$, and 
moreover, denoting by ${\mathfrak{U}}$ the universal enveloping algebra, ${\mathfrak{U}} ( Vir ) = {\mathfrak{U}} \, ( {\mathfrak{n}}_- ) \, 
{\mathfrak{U}} ( {\mathfrak{h}} ) \, {\mathfrak{U}} ( {\mathfrak{n}}_+ )$ (Poincar\'e-Birckhoff-Witt theorem), ${\cal{V}} = {\mathfrak{U}} ( Vir ) 
\, {\bf{v}} \equiv {\mathfrak{U}} ( {\mathfrak{n}}_- ) \, {\bf{v}}$, and ${\mathfrak{n}}_+ \, {\bf{v}} = 0$. In particular, a highest weight 
representation of $Vir$ in ${\mathfrak{U}} ( Vir ) [ C , h]$ with highest weight vector ${\bf{v}}$ and weights $( C , h )$ is a Verma 
representation \cite{Kac} if it satisfies a further constraint of universality, namely that any two such highest weight representations are 
connected to each other by a unique epimorphism. This is equivalent to saying that there exists at most one highest weight representation of 
$Vir$ for given weights $( C , h )$, which is the Verma representation. The latter is unitary iff $C \geq 0$ and $h \geq 0$.     
  
In two--dimensional conformal field theory and vertex operator algebra theory, modular functions and modular forms appear often as graded dimensions, 
or characters, of infinite dimensional irreducible modules. Although individual characters are not always modular in this way, it can be the case that the 
vector spaces spanned by all of the irreducible characters of a module are invariant under the modular group. In the case of the Virasoro 
vertex operator algebras, the Wronskians of a basis of the module were thoroughly studied \cite{Mil}, leading to several classical $q$--series 
identities related to modular forms, using methods from representation theory. Of particular interest here is a class of Virasoro algebras 
related in conformal field theory to the so called Virasoro minimal models ${\cal{M}} (2, 2k+1)$ ($k>2$ an integer) \cite{Mil}, \cite{FeFr}, 
\cite{KaWa}, \cite{RoCa}. For $1\leq i \leq k$, let  
\begin{equation}
c_k \doteq -2 \, \frac{(k-1)(6k-1)}{2k+1} \quad , \quad h_{i,k} \doteq - \frac{1}{2} \, \frac{(2k-i)(i-1)}{2k+1} \; , \nonumber 
\end{equation} 
and let $\Lambda ( c_k , h_{i,k} )$ denote the (irreducible) lowest weight module for the Virasoro algebra with central charge $c_k$ and 
weight $h_{i,k}$. The representations are ${\mathbb{N}}$-gradable and have finite-dimensional graded subspaces. One can therefore define 
the formal $q$--series 
\begin{equation}
{\rm{dim}}_{\Lambda (c_k , h_{i,k})} (q) \doteq \sum_{n \in {\mathbb{N}}} {\rm{dim}} \left ( \Lambda (c_k , h_{i,k})_n \right ) \, q^n \; , 
\nonumber
\end{equation}
The expression   
\begin{equation} 
{\rm{ch}}_{i,k} (q) \equiv q^{\, h_{i,k}-\frac{c_k}{24}}\, {\rm{dim}}_{\Lambda (c_k , h_{i,k})} (q) \; , \nonumber 
\end{equation}
is called the $'$character$\, '$ of $\Lambda ( c_k , h_{i,k} )$. It turns out \cite{KaWa}, \cite{RoCa}, that  
\begin{equation}
{\rm{ch}}_{i,k} (q) = q^{\, h_{i,k}-\frac{c_k}{24}} \cdot \prod_{{n \in {\mathbb{N}}}\atop{n \neq \, 0 , \pm i\, ({\rm{mod}} 2k+1)}} 
\frac{1}{1-q^n} \; . \nonumber 
\end{equation}  
One defines then the Wronskian in the following way. Let $W_k (q)$ and $W_k' (q)$ be $k \times k$ matrices of the form  
\begin{equation}
W_k (q) \doteq \left | \begin{array}{c c c c} {\rm{ch}}_{1,k} (q) & {\rm{ch}}_{2,k} (q) & \cdots & {\rm{ch}}_{k,k} (q) \\ 
{\rm{ch}}_{1,k}' (q) & {\rm{ch}}_{2,k}' (q) & \cdots & {\rm{ch}}_{k,k}' (q) \\
\vdots & \vdots & \vdots & \vdots \\ 
{\rm{ch}}_{1,k}^{(k-1)} (q) & {\rm{ch}}_{2,k}^{(k-1)} (q) & \cdots & {\rm{ch}}_{k,k}^{(k-1)} (q) \\
\end{array} \right | \; , \nonumber 
\end{equation}
\begin{equation}  
W_k' (q) \doteq \left | \begin{array}{c c c c} {\rm{ch}}_{1,k}' (q) & {\rm{ch}}_{2,k}' (q) & \cdots & {\rm{ch}}_{k,k}' (q) \\ 
{\rm{ch}}_{1,k}^{(2)} (q) & {\rm{ch}}_{2,k}^{(2)} (q) & \cdots & {\rm{ch}}_{k,k}^{(2)} (q) \\
\vdots & \vdots & \vdots & \vdots \\ 
{\rm{ch}}_{1,k}^{(k)} (q) & {\rm{ch}}_{2,k}^{(k)} (q) & \cdots & {\rm{ch}}_{k,k}^{(k)} (q) \\
\end{array} \right | \; , \nonumber  
\end{equation}
where ${\rm{ch}}_{m,k}^{(\ell )} (q) = \bigl ( {\rm{ch}}_{m,k}^{(\ell -1)} (q) \bigr )'$, $\ell \geq 2$, ${\rm{ch}}_{m,k}^{(1)} (q) \equiv 
{\rm{ch}}_{m,k}' (q)$, and the derivation denoted by a prime is defined as $\displaystyle{\left ( \sum_n a(n) q^n \right )' \doteq \sum_n a(n) 
n q^n}$. This definition is due to the fact that for $q = \exp ( i 2\pi z)$, it is equivalent to $\displaystyle{\frac{1}{i 2\pi}\, 
\frac{{\rm{d}}}{{\rm{d}}z}}$. 

Given the two matrices above, one defines the Wronskians ${\cal{W}}_k (q)$ and ${\cal{W}}_k' (q)$ by 
\begin{equation}
{\cal{W}}_k (q) \doteq \alpha (k) \, {\rm{det}} W_k (q) \; , \; {\cal{W}}_k' (q) \doteq \beta (k) \, {\rm{det}} W_k' (q) \; , \nonumber 
\end{equation}
where $\alpha (k)$ and $\beta (k)$ are rational numbers defined in such a way that the corresponding power series in $q$ has leading coefficient 
(whenever it is non-vanishing) equal to $1$. It was shown \cite{Dau} that ${\cal{W}}_k (q)$ is non-vanishing for all values of $k$ and, upon 
setting $\displaystyle{a_{i,k} \doteq h_{i,k} - \frac{c_k}{24}}$, the two scaling factors are given by: 
\begin{eqnarray}
\alpha (k) &=& \prod_{1\leq i < j\leq k} \left ( a_{j,k} - a_{i,k} \right )^{-1} \; , \nonumber \\  
\beta (k) &=& \Biggl \{ \begin{array}{l l} 0 & {\rm{if}} \; a_{i,k} = 0 \; {\rm{for}} \; {\rm{some}} \; 1 \leq i \leq k \; , \\ 
\alpha (k) \, {\displaystyle{\prod_{1\leq i \leq k} a_{i,k}^{-1}}}  & {\rm{otherwise}} \; . \\ \end{array} \nonumber 
\end{eqnarray}
By expanding the matrix elements of $W_k (q)$ and applying the differentiation operator one gets 
\begin{equation}
{\cal{W}}_k (q) = \alpha (k) \cdot {\rm{det}} \left | \begin{array}{c c c c} q^{a_{1,k}} + \dots & q^{a_{2,k}} + \dots & \cdots & q^{a_{k,k}} + 
\dots \\ a_{1,k} q^{a_{1,k}} + \dots & a_{2,k} q^{a_{2,k}} + \dots & \cdots & a_{k,k} q^{a_{k,k}} + \dots \\
\vdots & \vdots & \vdots & \vdots \\ 
a_{1,k}^{k-1} q^{a_{1,k}} + \dots & a_{2,k}^{k-1} q^{a_{2,k}} + \dots & \cdots & a_{k,k}^{k-1} q^{a_{k,k}} + \dots \\
\end{array} \right | \; . \nonumber 
\end{equation}
It is interesting to the purpose of this paper to notice that the smallest power of $q$ in the power series for ${\cal{W}}_k (q)$ can be calculated   
exclusively in terms of these leading values. Let $w_k$ be the leading coefficient of ${\rm{det}} W_k (q)$; clearly 
\begin{equation}
w_k = {\rm{det}} \left | \begin{array}{c c c c} 1 & 1 & \cdots & 1 \\
a_{1,k} & a_{2,k} & \cdots & a_{k,k} \\ 
\vdots & \vdots & \vdots & \vdots \\ 
a_{1,k}^{k-1} & a_{2,k}^{k-1} & \cdots & a_{k,k}^{k-1} \\ \end{array} \right | \; , \label{wk} 
\end{equation}   
Even though this was to be expected (and is indeed almost tautologic: if ${\rm{det}} W_k (q)$ $\neq 0$, then $w_k = \alpha (k)^{-1}$ by definition), 
it is however surprising and crucial that the above expression is a Vandermonde determinant with entries in ${\mathbb{Q}}\,$.  

The super (${\mathbb{Z}}_2$-graded) Virasoro algebra $sVir$ is an infinite $2$-graded algebra containing $Vir$ (with $c=0$) as bosonic sector 
subalgebra, with generators $\{ L_n , G_n , F_n \, | \, n \in  {\mathbb{Z}} \}$, satisfying the defining relations \cite{RASO} 
\begin{eqnarray}
& & \left [\!\left [ L_n , L_m \right ]\!\right ] = (n-m) L_{n+m} \quad , \quad \left [\!\left [ F_n , G_m \right ]\!\right ] = G_{n+m} 
\; , \nonumber \\ 
& & \left [\!\left [ L_n , F_m \right ]\!\right ] = - m F_{n+m} \quad , \quad \left [\!\left [ F_n , F_m \right ]\!\right ] = 0 
\; , \nonumber \\  
& & \left [\!\left [ L_n , G_m \right ]\!\right ] = (n-m) G_{n+m} \quad , \quad \left [\!\left [ G_n , G_m \right ]\!\right ] = 0 
\; , \nonumber 
\end{eqnarray}
where the notation was introduced ($'$super$\, '$ commutation bracket) 
\begin{equation} 
\left [\!\left [ X , Y \right ]\!\right ] \doteq X Y - (-)^{deg (X)\, deg (Y)}\, Y X \; , \nonumber 
\end{equation}
and the ${\mathbb{Z}}_2$-grading is achieved by requiring that 
\begin{equation}
deg (L_n) = 0 = deg (F_n) \; , \; ({\rm{bosonic}}) \; ; \; deg (G_n) =1 \; , \; ({\rm{fermionic}}) \; , \; 
\forall n \in {\mathbb{Z}} \; .  \nonumber 
\end{equation}
A realization of the graded algebra $sVir$ in terms of generators of $su(1,1)$ ($\{ K_+ , K_- , K_3 \}$, with commutation relations $[ K_+ , 
K_- ] = - 2 K_3$, $[ K_3 , K_{\pm} ] = \pm K_{\pm}$), and of the Clifford algebra fermionic generators $f_+$, $f_-$ (with anti-commutation 
relations $\{ f_+ , f_- \} = {\mathbb{I}}$, $\{ f_{\pm} , f_{\pm} \} = 0$), is given by  
\begin{eqnarray}
L_k &=& i \frac{1}{2^{k+1}}  \{ K_3^{-1} , K_{+} \}^{k} \{ K_3^{-1} , K_{-} \} \; , \nonumber \\ 
G_k &=& \frac{1}{2^{k+1}} \{ K_3^{-1} , K_{+} \}^k f_+ \{ K_3^{-1} , K_{-} \} \; , \nonumber \\ 
F_k &=& \{ K_3^{-1} , K_{+} \}^{k-1} f_+ f_- \; , \; k \geq 0 \; , \nonumber   
\end{eqnarray}
where the representation to be adopted for $su(1,1)$ is constrained by the additional hermiticity constraints $L_{-k} \doteq L_k^{\dagger}$, 
$G_{-k} \doteq , G_k^{\dagger}$, and $F_{-k} \doteq F_k^{\dagger}$. The bosonic sector Witt algebra in such realization corresponds to the 
algebra of vector fields over the unit circle $S^{(1)}$, $\displaystyle{z = {\rm{e}}}^{i \theta}$, with $\displaystyle{L_n = i {\rm{e}}}^{i 
n \theta}\, \frac{{\rm{d}}}{{\rm{d}} \theta}$ \cite{Rase}.  

The latter realization is not unique. An (inequivalent) one \cite{DoFa}, holding for $Vir$ (possibly extendable to $sVir$, but the result is not 
known yet), is given in terms of a bosonic Virasoro primary field $a_n$ of zero conformal weight and of its canonical conjugate $a_n^{\dagger}$, 
of weight 1, $n\in {\mathbb{Z}}$, 
\begin{equation}
\left [ a_m^{\dagger} , a_n \right ] = \delta_{m+n} \; , \; \left [ a_m^{\dagger} , a_n^{\dagger} \right ] = 0 = \left [ a_m , a_n \right ] 
\; , \nonumber  
\end{equation}
by 
\begin{equation}
L_n = \sum_{r \in {\mathbb{Z}}} r : a^{\dagger}_{n-r} a_r : + \lambda (n+1) \left ( a_n^{\dagger} + M n a_n \right ) \; , \nonumber 
\end{equation}
which implies
\begin{eqnarray}
\left [ L_m , a_n \right ] &=& (m+n) a_{m+n} - \lambda (m+1) \delta_{m,-n} \; \nonumber \\  
\left [ L_m , a_n^{\dagger} \right ] &=& n a_{m+n}^{\dagger} - \lambda m (m+1) M \delta_{m,-n} \; . \nonumber 
\end{eqnarray}
Here, as customary, ${\bf{:}} {\bf{:}}$ denotes normal ordering, and $\lambda$ and $M$ are free parameters connected with the central 
charge $c$; $c = 2 - 24\, M \lambda^2$. 

This realization has a fermionic counterpart, with a fermionic Virasoro primary field $f_r$ of weight $\lambda$ and its canonical conjugate 
$f_r^{\dagger}$, of weight $1-\lambda$, $r\in {\mathbb{Z}} + \frac{1}{2}$,  
\begin{equation}   
\{ f_r , f_s^{\dagger} \} = \delta_{r+s} \; , \; \{ f_r^{\dagger} , f_s^{\dagger} \} = 0 = \{ f_r , f_s \} \; , \nonumber 
\end{equation}
implemented through  
\begin{equation}
L_n = - \sum_{r \in {\mathbb{Z}} + \frac{1}{2}} (r -\lambda n) \, : f^{\dagger}_{n-r} f_r : \; , \nonumber 
\end{equation}
from which follow the commutation relations 
\begin{eqnarray} 
\left [ L_m , f_r \right ] = ((1-\lambda )m+r)\, f_{m+r} &,& \left [ L_m , f_r^{\dagger} \right ] = (\lambda m+r) f_{m+r}^{\dagger} \; . \nonumber 
\end{eqnarray}

\section{Logical operators}

\subsection{Generalized bit flip operator}

A set of logical operators $X^{(k)}$ were defined in \cite{RaRa}; 
\begin{equation}
X^{(k)} \, = \, E^{(k)} \, {\hat{n}}^{- \frac{1}{2}} \, a^{\dagger} \; , \nonumber 
\end{equation} 
where $E^{(k)}$ is given in terms of $k$-boson algebra's operators 
\begin{equation} 
E^{(k)} \, = \, \left ( {\mathbb{I}} + {\hat{N}}_k \right )^{- \frac{1}{2}} A_k \, F^{(k)} + G^{(k)} \; . \label{Ek}
\end{equation}

These operators act in a simple way on the orthonormal computational basis of a $k$-dimensional $'${\sl{qukit}}$\, '$ 
constructed in the following way. Let 
\begin{equation}
|\omega_k \rangle \, = \, \sum_{j=0}^{k-1} \, c_j \, |j\rangle \, = \, \sum_{j=0}^{k-1} \left( \prod_{\ell =0}^{j-1} 
\sin \varphi_{\ell} \right ) \gamma_j \, {\rm{e}}^{i \mu_j} \, | j \rangle \; , \nonumber   
\end{equation}
be the normalized highest weight vector of the $k$-boson algebra \cite{k-boson-alg}, annihilated by $A_k$ and parametrized by $(k-1)$ angles 
$\varphi_{\ell}$, $\ell = 0, \dots , k-2$ ($k \geq 2$), where for simplicity coefficients $c_j$ are assumed to be real, $\gamma_j \, = \, 1$ 
for $j = k-1$, $\gamma_j \, = \, \cos \varphi_j$ otherwise, $\displaystyle{\prod_{\ell} \sin \varphi_{\ell} \doteq 1}$ for $j = 0$. 

The coherent states 
\begin{equation}
| {\bar{j}} \rangle \doteq {\rm{e}}^{- \frac{1}{2} \, |\beta |^2} \, {\rm{e}}^{\beta A_k^{\dagger}} \, |\omega_k \rangle =  
{\rm{e}}^{- \frac{1}{2} \, |\beta |^2} \, \sum_{\ell=0}^{\infty} \frac{\beta^{\ell}}{\sqrt{\ell !}} \, |k \ell +j \rangle \; , 
\; j = 0,1,\dots ,k-1 \, , \label{CODE2}  
\end{equation}
are the codewords looked for. Naturally for $k = 2$ one retrieves the usual binary computational scheme: indeed in this case the states 
$| {\bar{0}} \rangle$, $| {\bar{1}} \rangle$ are the even and odd components, respectively, of the coherent states of the $2$-boson algebra 
and define the computational basis of a qubit.   

$X^{(k)}$ and ${X^{(k)}}^{\dagger}$ realize the following two basic $'$bit-flip$\, '$ operations on the codewords (\ref{CODE2})
\begin{equation}
X^{(k)} \, | {\bar{j}} \rangle \, = \, | {\overline{j+1}} \; ({\rm{mod}}\, k) \rangle \quad , \quad {X^{(k)}}^{\dagger} \, 
| {\bar{j}} \rangle \, = \, | {\overline{j-1}} \; ({\rm{mod}}\, k) \rangle \; . \nonumber
\end{equation}

The matrix representation of $X^{(k)}$ with respect to the computational basis $\{ |{\bar{j}} \rangle \, | \, j=0, 
\dots , k-1 \}$ (\underline{not} to the Fock number-states $|k\ell +j\rangle$, $\ell \geq 0$, $j=0,1, \dots ,k-1$), is given by the traceless unitary matrix 
$R^{(k)}$ of elements
\begin{eqnarray}
R^{(k)}_{1,k} \,=\, 1 \quad , \quad  R^{(k)}_{j,j-1} \,=\, 1 \;\, {\rm for} \;\, 2 \leq j \leq k \; , \nonumber \\ 
R^{(k)}_{j,m} \,=\, 0 \;\,  {\rm{for}} \;\, j \neq 1 \, {\textstyle{\&}} \, m \neq j-1 \; {\rm{and}} \; j=1 \, {\textstyle{\&}} \, m \neq k \; , \nonumber 
\end{eqnarray}
which for $k = 2$ reduces to the bit flip operator of the binary computational scheme. 

In (\ref{Ek}), operator $E^{(k)}$ consists of a purely combinatorial regular factor,  
\begin{equation}
F^{(k)} \,=\, \frac{(-)^{k-1}}{(k-1)!} \, \prod_{j=1}^{k-1} \left ( \hat{D}_k - j \right ) \,=\, \frac{(-)^{k-1}}{(k-1)!} \, 
\sum_{m=1}^{k} S_k ^{(m)} {\hat{D}}_k^{m-1} \; , \nonumber
\end{equation}
where the coefficients $S_k ^{(m)}$ are Stirling numbers of the first kind \cite{ABRAM}, and of a part taking into account 
the residues in the modular arithmetics we are overimposing on the Fock space, represented by the operator $G^{(k)}$, given 
by 
\begin{equation}
G^{(k)} \,=\, \sum_{\ell=1}^{k-1} p_{\ell}^{(k)} {\hat{D}}_k^{\ell} \; . \nonumber 
\end{equation}
The polynomial of degree $k-1$ (in the variable ${\hat{D}_k}$'s) at the {\sl{r.h.s.}} interpolates the $k$ points $(0,0), 
(1,1), (2,1),..., (k-1,1)$. Its coefficients $p_l^{(k)}$ are the solution of the system of algebraic equations 
\begin{equation}
{\mathfrak{M}} | {\mathfrak{p}}\rangle \,=\, |{\mathfrak{b}}\rangle \; , \nonumber  
\end{equation} 
where $| {\mathfrak{p}}\rangle , |{\mathfrak{b}}\rangle$ are $k$-component vectors, the former with components $p_l^{(k)}$, 
the latter with components $b_0 =0$, $b_{\ell} = 1$, $\ \leq \ell \leq k-1$, whereas the matrix 
\begin{equation}
{\mathfrak{M}} = \left | \begin{array}{c c c c c} 
1 & 0 & 0 & \cdots & 0 \\
1 & 1 & 1 & \cdots & 1 \\ 
1 & 2 & 2^2 & \cdots & 2^{k-1} \\
\vdots & \vdots & \vdots & \vdots & \vdots \\ 
1 & (k-1) & (k-1)^2 & \cdots & (k-1)^{k-1} \\
\end{array} \right | \; , \label{matrix-M} 
\end{equation}
is but a special case of Vandermonde matrix (indeed it is just the matrix entering $\Delta$ in (\ref{Delta}) and $w_k$ in (\ref{wk}), in which 
$N \equiv k$, and $z_j\equiv j$), therefore the coefficients $p_l^{(k)}$ are readily obtained by the well known inversion formula for Vandermonde 
matrices \cite{NEA}, \cite{SERRA}. 

Resorting, for polynomial (\ref{polyn}) (once more with $N \equiv k$), to the expansion (\ref{sym}) in terms of symmetric polynomials 
${\mathfrak{s}}_q (z_1, \cdots, z_k)$, sum of all the inequivalent products of degree $q$, $0 \leq q \leq k$, in the $k$ variables $z_1, \cdots, 
z_k$, ${\mathfrak{s}}_0 (z) \doteq 1$, and setting 
\begin{equation}
{\mathfrak{s}}^{[m]}_q \doteq {\mathfrak{s}}_q (z_1, \dots, z_{m-1}, z_{m+1}, \dots, z_k)\; {\rm{with}} \; z_r \equiv r-1, r = 1,\dots, 
k \; , \nonumber 
\end{equation} 
one has, with $p_0^{(k)} \equiv 0 \, ; \, \forall k$:  
\begin{equation}
p_l^{(k)} = \sum_{m=2}^{k} (-)^{l+1+m} \, {{\mathfrak{s}}^{[m-1]} _{k-1-l}} \, \prod_{{j=1}\atop{(j\neq m)}}^{k} |j - m|^{-1} \; 
, \; 1\leq l\leq k-1 \; . \nonumber
\end{equation}

This straightforwardly leads \cite{RasRaf} to have, as eigenvalue of the operator ${\hat{N}}_k$ (${\hat{n}} = k {\hat{N}}_k + {\hat{D}}_k$),    
\begin{equation}
\left \lfloor  \frac{n}{k} \right \rfloor = \frac{2n-k+1}{2k} + \sum_{j=1}^{k-1} C_j^{(k)} \, \zeta_j^n \; , \nonumber
\end{equation} 
where $\lfloor x \rfloor$ denotes the integral part of $x$, namely the maximum integer $\leq x$, $\displaystyle{\zeta_{\ell} = \exp \left ( i 
2\pi \frac{\ell}{k} \right )}$, $\ell = 1, \dots , k-1$, and, setting $\zeta_0 \doteq 1$,  
\begin{equation}
C_j^{(k)} = \left ( \zeta_j - 1 \right )^{-1} \, \prod_{{\ell =0}\atop{( \ell \neq j )}}^{k-1} \left ( \zeta_j - \zeta_{\ell} \right )^{-1}
\; , \; k \geq 2 \; . \nonumber 
\end{equation}
It is intriguing that $(\zeta_j -1)\, C_j^{(N)}$ equals the inverse of ${\mathfrak{D}} (\zeta_0 , \zeta_1 , \dots , \zeta_{N-1})$, as 
defined in (\ref{star}).  

The above realization of logical operators in terms of $k$-boson operators has interesting features with respect to an underlying group structure
that was introduced several years ago in the context of squeezed states \cite{KaRaSo}. Let ${\mathfrak{F}}_{(k)}$ denote the transformation,  
${\mathfrak{F}}_{(k)} \, : \, h(1) \mapsto h(1)$, whereby $k$-boson operators ($k\in {\mathbb{N}}\, ;\, k\geq 1$) are constructed out of creation 
and annihilation operators for ordinary bosons; ${\mathfrak{F}}_{(k)} \, : \, a^{\dagger} \mapsto A_k^{\dagger}$ (or $A_k^{\dagger} = 
{\mathfrak{F}}_{(k)} (a^{\dagger})$, where $\displaystyle{{\mathfrak{F}}_{(k)} (a) \equiv \left [ {\mathfrak{F}}_{(k)} (a^{\dagger}) \right 
]^{\dagger}}$ and ${\mathfrak{F}}_{(1)} (a^{\dagger}) = a^{\dagger}$, ${\mathfrak{F}}_{(0)} = \emptyset$). For given $k$, the non-linear 
transformations ${\mathfrak{F}}_{(k)}$ generate a semigroup. In view of the property that, for $k , \ell \in {\mathbb{Z}}$,  
\begin{equation}
{\mathfrak{F}}_{(k)} \circ {\mathfrak{F}}_{(\ell )} (a^{\dagger}) = {\mathfrak{F}}_{(k \ell )} (a^{\dagger}) \; , \nonumber  
\end{equation}   
such semigroup can be extended formally to the abelian group of canonical transformations over pairs of conjugate operators (that for simplicity 
we denote by the same symbol), 
\begin{equation} 
\left \{ {\mathfrak{F}}_{(r)} \, | \, r \in {\mathbb{Q}} \, , \, r>0 \right \} \; . \nonumber  
\end{equation}
Indeed, one can define first the inverse transformation ${\mathfrak{F}}_{(k)}^{-1}$, through 
\begin{equation}
{\mathfrak{F}}_{(k)}^{-1} \circ {\mathfrak{F}}_{(k)} (a^{\dagger}) = a^{\dagger} = {\mathfrak{F}}_{(1)} (a^{\dagger}) \; , \nonumber 
\end{equation}
and -- upon equating ${\mathfrak{F}}_{(k)}^{-1} = {\mathfrak{F}}_{(1/k)}$ -- obtain eventually, for $r = \ell / k$, $r\in {\mathbb{Q}}$,  
\begin{equation}
{\mathfrak{F}}_{(k)}^{-1} \circ {\mathfrak{F}}_{(\ell )} = {\mathfrak{F}}_{(\ell /k)} \equiv {\mathfrak{F}}_{(r)} \; . \nonumber 
\end{equation}
${\mathfrak{F}}_{(r)}$ was assumed to generate what were referred to in \cite{KaRaSo} as $''$fractionary bosons$\, ''$. 

We argue that, on the one hand these correctly describe the collective excitations of the FQHE, whereas on the other the construction reported 
should allow us to implement the logical operators $X^{(k)}$ in terms of anyons, namely as $X^{(r)}$, generalizing the notion of $'$qukit$\, '$ 
to that of {\sl{qurit}}, with $r$ rational.

\section{The quantum field theoretical perspective}

\subsection{Chern-Simons matrix models and Laughlin wave-functions}
Another unexpected connection comes from the deep relation that exists between FQHE and non-commutative quantum field theory, pointed out by 
Susskind \cite{SUS}, who proved how Laughlin states at filling fraction $\nu$ for a system of (infinitely many) electrons in the lowest 
Landau level could be described by a $U(1)$ Chern-Simons theory.  

The action that describes the matrix Chern-Simons model is given by 
\begin{equation} 
S = \frac{1}{2} B\, \int {\rm{d}}t \left [ {\rm{Tr}} \left \{ \epsilon_{a, b} \left ( {\dot{X}}_a + i \left [ A_0 , X_a \right ] \right ) X_b 
\right \} + \Psi^{\dagger} \left ( i {\dot{\Psi}} -A_0 \Psi \right ) \right ] \; , \label{Action} 
\end{equation}
where $X_a$, $a=1,2$, are $N\times N$ matrices and $\Psi$ is a complex $N$-vector, that transform in the natural way under the action of the 
gauge group $U(N)$, 
\begin{equation}
X_a \rightarrow U X_a U^{-1} \quad , \quad \Psi \rightarrow U \Psi \; . \nonumber  
\end{equation} 
The equation of motion for $A_0$ induced by such action gives rise to the constraint 
\begin{equation}
{\cal{C}} \doteq -iB \, \left [ X_1 , X_2 \right ] + \Psi \Psi^{\dagger} - B \theta = 0 \; , \nonumber 
\end{equation}
where $\theta$ is given by the equation  
\begin{equation}
\Psi^{\dagger} \Psi = NB\theta \; , \nonumber 
\end{equation}
implied taking the trace of ${\cal{C}}$. 

Notice that if $\Psi$ is not present, parameter $\theta$ must vanish: in this case the action $S$ becomes that of a one-dimensional hermitian 
matrix model in an harmonic potential, which has been shown to be equivalent to a system of $N$ fermions in a 
confining potential \cite{PAR}. 
Quantization leads then to commutation relations that are the obvious generalization of (\ref{commut}): 
\begin{equation}
\left [ \Psi_k , \Psi_j \right ] = \delta_{k,j} \quad , \quad \left [ \left ( X_1 \right )_{\ell ,j} , \left ( X_2 \right )_{k ,m} \right ] = 
\frac{i}{B} \, \delta_{\ell ,m} \, \delta_{j,k} \; , \nonumber 
\end{equation} 
with an Hamiltonian which describes a set of $N(N-1)$ oscillators coupled via the constraint ${\cal{C}}$, of the form 
\begin{equation}
H = \omega \left ( \frac{1}{2} N^2 + \sum_{j,k =1}^N A_{\ell ,k}^{\dagger} A_{k,\ell} \right ) \; , \; A \doteq \sqrt{\frac{1}{2}B} \, 
\left ( X_1 + i X_2 \right ) \; . \nonumber  
\end{equation}
Indeed, one can check that, upon quantization, ${\cal{C}}$ becomes the generator of unitary transformations of both $X_a$ and $\Psi$, $NB\theta$ 
must be an integer, whence $B \theta = \kappa$, with $\kappa$ integer, and the physical states are singlets of $SU(N)$. In particular, the ground 
state must be of the form 
\begin{equation}
|\Psi_{\kappa}\rangle = \left [ \epsilon^{i_1 \cdots i_N} \Psi_{i_1}^{\dagger} \bigl ( \Psi^{\dagger} A^{\dagger} \bigr )_{i_2} \cdots 
\bigl ( \Psi^{\dagger} {A^{\dagger}}^{N-1} \bigr )_{i_N} \right ]^{\kappa} |0\rangle \; , \label{ground} 
\end{equation} 
where the vacuum $|0\rangle$ is annihilated by both $A$'s and $\Psi$'s. 

On this structure coherent states $|z,\psi\rangle$ can be defined, 
\begin{equation} 
A_{m,n} |z,\psi\rangle = z_{m,n} |z,\psi\rangle \quad , \quad \Psi_n |z,\psi\rangle = \psi_n |z,\psi\rangle \; , \nonumber 
\end{equation}
where, of course, $z$ is now a $N\times N$ matrix, and $\psi$ an $N$-component complex vector. 

We can express the norm of the ground state (\ref{ground}) in terms of these coherent states, resorting to their completeness, 
\begin{equation} 
\langle\Psi_{\kappa}|\Psi_{\kappa}\rangle = \int \prod_{k,j,\ell =1}^N {\rm{d}}{\bar{z}}_{k,j} \, {\rm{d}}z_{k,j} \, {\rm{d}}{\bar{\psi}}_{\ell} 
\, {\rm{d}} \psi_{\ell}\, \langle\Psi_{\kappa}|z,\psi\rangle \langle z,\psi |\Psi_{\kappa}\rangle \; . \nonumber    
\end{equation}
The latter can be straightforwardly turned into the form, analogous to the quantum mechanical QFHE case,    
\begin{equation} 
\langle\Psi_{\kappa}|\Psi_{\kappa}\rangle = \int \prod_{k,j,\ell =1}^N {\rm{d}}{\bar{z}}_{k,j} \, {\rm{d}}z_{k,j} \, {\rm{d}}{\bar{\psi}}_{\ell} 
\, {\rm{d}} \psi_{\ell}\, \prod_{{k,j=1}\atop{k<j}}^N |z_k - z_j|^{2\kappa} \, {\rm{e}}^{- {\bar{\psi}} \psi} \; . \nonumber 
\end{equation}
 
Upon parametrizing the coordinates of the $N \rightarrow \infty$ electrons in an external magnetic field in a $''$fuzzy$\, ''$ (mean field) way 
in terms of two (now infinite) hermitian matrices $X_a$, $a=1,2$, (units such that electric charge $q=1$), gives rise to a gauge action of the 
form \cite{SUS}
\begin{equation} 
S = \frac{1}{2} B\, \int {\rm{d}}t {\rm{Tr}} \left \{ \epsilon_{a, b} \left ( {\dot{X}}_a + i \left [ A_0 , X_a \right ] \right ) X_b 
+ 2 \theta A_0 \right \} \; , \label{Action-1} 
\end{equation}
where the inverse of parameter $\theta$ can now be explicitly recognized to measure the electron density $\varrho$ (indeed $\varrho = (2\pi 
\theta )^{-1}$). 

If one identifies the $'$coordinates$\, '$ $X_a$ with the (covariant) derivative operator, $X_a \sim \theta D_a$, and sets $D_0 = -i 
\partial_t + A_0$, such action turns out to have the form of a non-commutative Chern-Simons action \cite{Poly}, whose well known operatorial 
form is 
\begin{equation}
S = \frac{B \theta}{4 \pi} \int {\rm{d}}t \, 2\pi\theta \, {\rm{Tr}} \left \{ \frac{2}{3}\, \epsilon_{\mu ,\nu ,\rho} \, D_{\mu} D_{\nu} D_{\rho} 
+ \frac{2}{\theta}\, A_0 \right \} \; . \nonumber 
\end{equation}
The time component $A_0$ of the gauge field ensures here gauge invariance; its equation of motion indeed imposes the Gauss law constraint, 
which in turn amounts to 
\begin{equation}
\left [ X_1 , X_2 \right ] = i \theta \; . \nonumber 
\end{equation}

Gauge transformations are conjugations of $X_a$ or $D_a$ by arbitrary (possibly time-dependent) unitary operators; in the QHE context 
they have the meaning of reshuffling of the electrons. The generator of gauge transformations is 
\begin{equation}
G \doteq - i B \, \left [ X_1 , X_2 \right ] = B \theta \; , \nonumber 
\end{equation} 
which allows us to recognize $(B \theta )^{-1}$ as the filling fraction $\nu$. As gauge transformations are identified with reshuffling of 
particles, the above equation for $G$ has the interpretation of endowing the particles with quantum statistics of order $\nu^{-1}$. 

Coming finally to finite quantum Hall states consisting of $N$ electrons, obviously the $X_a$'s have to be represented by $N\times N$ matrices, 
for which the action (\ref{Action-1}) and constraint equation (\ref{commut}) are no longer consistent. A modified action has to be written 
to capture the true physical features of the FQHE, which differs from (\ref{Action}) simply for the terms taking into account fractionary 
filling and harmonic confinement: 
\begin{eqnarray} 
S &=& \frac{1}{2} B \int {\rm{d}}t \biggl [ {\rm{Tr}} \left \{ \epsilon_{a, b} \left ( {\dot{X}}_a + i \left [ A_0 , X_a \right ] \right ) X_b 
+ 2\theta A_0 - \omega X_a^2
\right \} \nonumber \\ &+& \Psi^{\dagger} \left ( i {\dot{\Psi}} - A_0 \Psi \right ) \biggr ] \; , \nonumber 
\end{eqnarray}
that interpolates between (\ref{Action}) and (\ref{Action-1}).    

It is intriguing that a formal structure which, based on the underlying combinatorial skeleton inherent in the logical operators for quantum 
information processing in terms of $k$--bosons, was constructed building over the physical platform of systems endowed with anyonic statistics, 
carried through a set of inequivalent but isomorphic representations, eventually made us land on the object (a discretized version of the 
Chern--Simons topological quantum field theory) that is the pillar over which the automaton--based quantum algorithm allowing us to estimate 
efficiently the invariants of 3--manifolds is rooted \cite{GaMaRa}, \cite{GaMaRa3}.     

\section{Conclusions}

In this essay we tried to point out the deep formal structural unity as well as the unexpected physical correlations among different quantum 
physical systems, and to explore how such correlations bear on the capacity of the system to carry (encode) and manipulate information. 
The role of logical operators constructed out of multi-boson (and possibly fractionary boson) operators was the starting point of the analysis, 
which has led us to identify its relation -- through the crucial ingredient of Vandermonde determinants, which embody all the combinatorial and 
transformational features of the theories considered -- on the one hand with the algebra of diffeomorphisms of the circle (crucially entering 
several fields of physics, mainly conformal field theories), on the other with the physical properties of the fractionary quantum Hall effect 
(as described in terms of Laughlin states) and of the many electron representation in terms of the Chern-Simons topological quantum field 
theory. The cross breeding among such a priori quite distant domain of theoretical physics promises to be fruitful of novel, unexpected, and -- 
if successful -- certainly far reaching results.


\begin{thebibliography}{99}
\bibitem{k-boson-alg}
What is here referred to as $k$-boson algebra is the algebra generated by $\{ A_k , A_k^{\dagger} , {\hat{n}} , {\mathbb{I}} \}$, where 
${\hat{n}} \equiv a^{\dagger} a$ is the customary number operator in Fock space, and $[ A_k , {\hat{n}} ] = k A_k$, $[ A_k, A_k^{\dagger} ] 
= {\mathbb{I}}$. In its enveloping algebra one defines the $k$-boson number $N_k \doteq A_k^{\dagger} A_k$, and the residue (mod $k$) of 
${\hat{n}}$, $D_k \doteq {\hat{n}} - k\, {\hat{N}}_k$. $\{ A_k , A_k^{\dagger} , {\hat{N}}_k , {\mathbb{I}} \}$ generate $h(1)$, because   
$[ A_k , {\hat{N}}_k ] = A_k$.
\end{thebibliography}
\end{document}